\documentclass[final,10pt]{elsart}

\usepackage{amssymb,latexsym,amsmath,amsfonts}
  \usepackage[dvips]{graphicx}
\usepackage{subfigure}
\usepackage{hyperref}

\usepackage{pdfpages}
\usepackage{dcolumn}
\usepackage{bm}
\usepackage{graphicx,subfigure,xcolor}

\usepackage{amsmath}
\usepackage{amssymb}
\usepackage{amsfonts}
\usepackage{latexsym}

\newcommand{\Hil}{\mathcal H}
\newcommand{\Dbos}{{\mathcal D}}
\newcommand{\F}{{\cal F}}

\newcommand{\1}{1 \!\! 1}

\begin{document}

\begin{frontmatter}

\title{Exceptional Points in a non-Hermitian extension of the Jaynes-Cummings Hamiltonian}
\author[autore,autore2]{Fabio Bagarello}
\author[autore]{Francesco Gargano}
\author[autore]{Margherita Lattuca}
\author[autore3]{Roberto Passante}
\author[autore3]{Lucia Rizzuto}
\author[autore3]{Salvatore Spagnolo}

\journal{}
\maketitle
\date{}
\address[autore]{D.E.I.M, University of Palermo, Italy.}
\address[autore2]{I.N.F.N, Torino, Italy.}
\address[autore3]{ Dipartimento di Fisica e Chimica, University of Palermo, Italy, Italy}
\begin{abstract}
\noindent We consider a generalization of the non-Hermitian ${\mathcal PT}$ symmetric Jaynes-Cummings {Hamiltonian, recently introduced for studying optical phenomena with time-dependent physical parameters, that includes environment-induced decay}. In particular, we investigate the interaction of a two-level fermionic system (such as a two-level atom) with a single bosonic field mode in a cavity. The states of {the} two-level system are allowed to decay because of the interaction with the environment, {and this is included phenomenologically in our non-Hermitian Hamiltonian by introducing complex energies for the fermion system. We focus our attention} on the occurrence of exceptional points in the spectrum of the Hamiltonian, clarifying its mathematical and physical meaning.
\end{abstract}

\end{frontmatter}

\section{Introduction}
\label{sec:Introduction}

{Quantum systems whose time evolution can be described by effective non-Hermitian Hamiltonians have been considered since a long time, for example in the framework of irreversible statistical mechanics or for describing decaying unstable systems \cite{SH09}.
Originally introduced to describe phenomenologically these important physical systems, non-Hermitian Hamiltonians have been initially used overlooking the well-known contradictions related to their compatibility with the basic principles of quantum mechanics \cite{B63}}.

{In recent years, it has been considered in the literature the possibility to describe realistic physical systems using non-Hermitian Hamiltonians whose eigenvalues are real} \cite{Mos10,BB98,MosPra09,BenBerMan02}.
 {This is mathematically meaningful because requiring Hermiticity is a sufficient but not necessary condition to have a real spectrum and a unitary time evolution.}
In fact, recently, it has been shown that non-Hermitian Hamiltonians with ${\mathcal PT}$ (Parity-Time) symmetry can have a real eigenvalue spectrum \cite{BB98,BBJ02,Bender07}. The same happens for non-Hermitian but pseudo-symmetric Hamiltonians \cite{mosta2002}, where the ${\mathcal PT}$ symmetry is replaced by a more abstract condition.
 {This new approach has produced several important results in the theory of quantum open systems, quantum optics, balanced gain-loss systems, for example}, both from theoretical and experimental point of view (see for example \cite{Zhen15} and references therein).

A key point of this topic is to understand what happens when a  ${\mathcal PT}$ symmetry breaking occurs in a Hamiltonian describing a physical system. As recently investigated (see for example \cite{Heiss04}), this may for example happen when one or more  {physical parameters in the Hamiltonian assume} specific values in the complex plane.
In more general terms, this aspect is linked to a wider problem addressed in non-Hermitian operator theory, that is the theory of exceptional points (EPs), term introduced in the literature by Kato \cite{Kato66}.
Many physical problems are described by Hamiltonians $H(\lambda)$ which manifest dependence on a parameter $\lambda$ linked to quantities accessible in the experimental setting.
Generally, the spectrum $E_n(\lambda)$ and eigenfunctions $|\psi_n (\lambda) \rangle$ of $H(\lambda)$  are analytic functions of $\lambda$.
It can occur that, for specific complex values of $\lambda$, two or more energy levels are equal and the corresponding eigenstates coalesce into a single state.  {It should be emphasized that, in the case of non-Hermitian Hamiltonians, this condition is very different from that of degeneracy common in quantum mechanics. In the presence of EPs, in fact, the coalescence of eigenstates causes the {collapse} of the subspace dimension to one.}
Because of this circumstance, the eigenstates no longer form a complete basis and this has very intriguing consequences. For example,  ${\mathcal PT}$ symmetry is broken \cite{Rott15}.
Moreover, EPs may play  {a very important role in several physical systems, for example in a photonic crystal slab where their presence has been shown to be connected with peaks of reflectivity \cite{Zhen15}.}

{In this paper, we shall consider a non-Hermitian generalization of the well-known Jaynes-Cummings (JC) Hamiltonian and investigate the occurrence of EPs in its spectrum. The Jaynes-Cummings model describes a two-level atom interacting with a mode of the quantized electromagnetic field in a cavity \cite{JC}; it has been extensively investigated in the literature, in particular in quantum optics. Very recently}, we have generalized  {this model} to the non-Hermitian {but PT-symmetric} case, in order to simulate a time-dependent modulation of the frequency of the two-level-atom {or} of the cavity mode in the presence of gain-loss \cite{BLPRS15}. We have also expressed the effective non-Hermitian Jaynes-Cummings Hamiltonian, having an imaginary coupling constant, frequency in term of pseudo-bosons and pseudo-fermions \cite{bagbook,BL13}, discussing also relevant mathematical and physical aspects of this extension of the Hamiltonian \cite{BLPRS15}.

 In this paper we focus our analysis on the occurrence of EPs in the spectrum of this extended Jaynes-Cummings Hamiltonian when the decay of the atomic states is allowed due to interaction with the environment, highlighting the main effects they have on the behavior of the system.
 {In order to make more general our analysis, we have modified the part of the Hamiltonian relative to the two-level system taking as a basis the analysis done in \cite{gmm}, where the authors} study nonadiabatic couplings in decaying systems, showing that EPs can influence time-asymmetric quantum-state-exchange mechanism.

The role played by the Jaynes-Cummings model has been crucial to the development of quantum optics and cavity electrodynamics, from both theoretical and experimental points of view \cite{CPP95}.  {Thus, our extension of the model can shed further light on the dynamics of open optical systems, usually described} in terms of non-Hermitian Hamiltonian. {Our analysis to elucidate the structure of the exceptional points of the spectrum of the deformed Jaynes-Cummings non-Hermitian Hamiltonian, can be relevant to understand the role of these points in the dynamics of physical systems, such as optical systems, that can be realized in the laboratory. Also, our analysis widens} the scenario of applications of the pseudo-bosons and pseudo-fermions formalism.

This paper is organized as follows.
 {In Sec. \ref{sec:nonHJC} we introduce our generalized Jaynes-Cummings model allowing the decay of the atomic states; in Sec. \ref{sec:EigH} we calculate exactly {the} spectrum and {the} eigenstates of $H$; in Sec. \ref{sec:EPform}, we discuss the formation of exceptional points in the extended Jaynes-Cummings model; Sec. \ref{sec:Concl} is dedicated to the discussions of our results and to our conclusive remarks.}

\section{The non-Hermitian Jaynes-Cummings Hamiltonian}
\label{sec:nonHJC}
 {The non-Hermitian extension of the Jaynes-Cummings Hamiltonian we are considering, written in terms of pseudo-bosons and pseudo-fermions, is}
\begin{equation}
H=H_{GMM}+\hbar\omega D d+\epsilon d C+\epsilon^* D c.
\label{31}
\end{equation}
 {where $\omega$ is the frequency of the boson field (a single cavity mode, for example), $\epsilon$ is the boson-fermion coupling constant; $c,C$ and $d,D$ are respectively the pseudo-bosons and pseudo-fermions satisfying the following commutation and anticommutation rules \cite{bagbook,BL13}}
\begin{eqnarray}
[d\otimes\1_f,D\otimes\1_f]=\1_b\otimes\1_f=\1 \, , \nonumber \\
\qquad \{\1_b\otimes c,\1_b\otimes C\}=\1 \, ,
\label{CommRules}
\end{eqnarray}
while all the other commutators are zero.
Our operators act on the Hilbert space $\Hil :=\Hil_b\otimes\Hil_f$, where $\Hil_f={\Bbb C}^2$ (fermionic sector) {and} $\Hil_b$ is infinite dimensional (bosonic sector); $\epsilon$ indicates the coupling constant.
The term $H_{GMM}$ in (\ref{31}) (GMM stands for Gilary, Mailybaev and Moiseyev), originally introduced in \cite{gmm} and studied later in \cite{FBgarg}, has the following form:
\begin{eqnarray}
H_{GMM}=\left(
  \begin{array}{cc}
    \epsilon_1-i\Gamma_1 & \nu_0 \\
    \nu_0 &  \epsilon_2-i\Gamma_2 \\
  \end{array}
\right),
\end{eqnarray}
where $\Gamma_1$ and $\Gamma_2$ are positive quantities, $\epsilon_1$ and $\epsilon_2$ are {real quantities}, and $\nu_0$ is complex-valued.
 {This term is an extension of the usual atomic term of the Jaynes-Cummings Hamiltonian and includes possibility of {decay} of the two atomic states to other states, for example due to the coupling with an environment with a continuous energy spectrum \cite{SH09}. The quantities $\Gamma_1$ and $\Gamma_2$ {can} be related to such phenomenological decay rates.}

As shown in \cite{FBgarg}, $H_{GMM}$ admits a (double) pseudo-fermions representation, and it can be written
as $H_{GMM}=\hbar\omega_0 N_f+\rho\1$ where $N_f=Cc$ and
\begin{equation}
c=\left(
       \begin{array}{cc}
        \alpha_{11} & \alpha_{12} \\
         -\alpha_{11}^2/\alpha_{12} & -\alpha_{11} \\
       \end{array}
     \right), \quad C=\left(
       \begin{array}{cc}
        \beta_{11} & \beta_{12} \\
         -\beta_{11}^2/\beta_{12} & -\beta_{11} \\
       \end{array}
     \right),
\label{add1}\end{equation}
\begin{eqnarray}\left\{
\begin{array}{ll}
\hbar\omega^{\pm}_0\gamma_{\pm}=\nu_0, \\
\alpha_{\pm}=\frac{1}{2\nu_0}\left(-\Delta\epsilon+i\Delta\Gamma\mp\sqrt{(-\Delta\epsilon+i\Delta\Gamma)^2+4\nu_0^2}\right),\\
\beta_{\pm}=\frac{1}{2\nu_0}\left(-\Delta\epsilon+i\Delta\Gamma\pm\sqrt{(-\Delta\epsilon+i\Delta\Gamma)^2+4\nu_0^2}\right),\\
\rho_{\pm}=\frac{1}{2}\left(\tilde\epsilon-i\Gamma\pm \sqrt{(-\Delta\epsilon+i\Delta\Gamma)^2+4\nu_0^2}\right),
\end{array}
\right.
\label{PFparameter}
\end{eqnarray}
being
$\alpha=\frac{\alpha_{11}}{\alpha_{12}}$, $\beta=\frac{\beta_{11}}{\beta_{12}}$, $\gamma_{\pm}=\alpha_{12}\beta_{11}-\alpha_{11}\beta_{12}=\alpha_{12}\beta_{12}(\beta_{\pm}-\alpha_{\pm})$, $\Delta\epsilon=\epsilon_2-\epsilon_1$, $\Delta\Gamma=\Gamma_2-\Gamma_1$, $\tilde\epsilon=\epsilon_2+\epsilon_1$ and $\Gamma=\Gamma_2+\Gamma_1$. We see that we have two possible solutions (the {\em plus} solution and the {\em minus} solution) and that both solutions admit two free parameters. For instance, choosing $\alpha_+$ above, implies that $\alpha_{11}$ is fixed if we first fix $\alpha_{12}$. Hence, for each given choice of $\alpha_{12}$ we have a different solution for $c$.
{It should be noted that the following condition for the pseudo-fermions existence} (see \cite{FBgarg}),
\begin{equation} -\gamma_{\pm}^2=\alpha_{12}\beta_{12},
\label{PFexistence}
\end{equation}
must be satisfied, otherwise $H_{GMM}$ can be expressed as standard fermionic operator ($C=c^{\dag}$) or it cannot be diagonalized.
Since $\gamma_{\pm}=\alpha_{12}\beta_{12}(\beta_{\pm}-\alpha_{\pm})$ and $\gamma_{\pm}^2=-\alpha_{12}\beta_{12}$, we find that,  {whichever $\alpha_{\pm}\neq\beta_{\pm}$, by taking}
$$
\alpha_{12}\beta_{12}=\frac{-\nu_0^2}{(-\Delta\epsilon+i\Delta\Gamma)^2+4\nu_0^2},
$$
the condition (\ref{PFexistence}) is satisfied.  {On the other hand, this is not possible if $\alpha_{\pm}=\beta_{\pm}$, that is $(-\Delta\epsilon+i\Delta\Gamma)^2=-4\nu_0^2$;} in this case the eigenvalues of  $H_{GMM}$ coalesce to the value $\frac{1}{2}(\tilde\epsilon -i\Gamma)$.
Notice that all the above conditions lead to the link of $\omega_0$ with the parameters defining $H_{GMM}$, as the following condition must be satisfied to ensure that  \eqref{PFexistence} is valid:

\begin{equation}
\hbar\omega_0^{\pm}=\pm\sqrt{4\nu_0^2+(-\Delta \epsilon+i\Delta \Gamma)^2}.\label{relomega0}
\end{equation}
Notice that this in general means that $\omega_0^\pm$ are complex quantities. To simplify the treatment we restrict here to the principal square root.
This relation will prove to be very important in the following {because it} highlights that the formation of {EPs} depends on the phenomenological parameters $\Gamma_1$ and $\Gamma_2$, justifying their introduction in Hamiltonian \eqref{31}.

\section{ {Eigenstates and eigenvalues of $H$}}
\label{sec:EigH}

With a simple extension of the procedure discussed in  \cite{BLPRS15} (see also \cite{CPP95})
we can rewrite $H$  in a diagonal
form. For that, we first introduce a global non self-adjoint
number operator, analogous to the total-excitations-number operator,
$$
N=Dd+Cc,
$$
and the map defined as
\begin{equation}
T=\textrm{exp}\{-\theta(4|\epsilon |^2N)^{-1/2}(\epsilon dC+ \epsilon^\ast Dc) \},
\end{equation}
where $\theta$ is the operator defined by $\sin\theta=-(4|\epsilon|^2N)^{1/2}\Delta^{-1}$ and
$\cos\theta=-\delta\Delta^{-1}$, where $\delta=\hbar(\omega_0- \omega)$ is the detuning between the energies of the two {fields} and $\Delta=\left(\delta^2+4|\epsilon|^2N\right)^{1/2}$.
By defining the  {\emph{dressed}} operators $\hat C=TCT^{-1}, \hat c=TcT^{-1}, \hat D=TDT^{-1}, \hat d=TdT^{-1}$
it {is} easy to check that  they are tensor products of pseudo-bosonic and pseudo-fermionic operators satisfying themselves  commutation rules analogous to (\ref{CommRules}), and the Hamiltonian $H$ can be written in a diagonal form as:
\begin{equation}
H=\left(\hbar\omega- \hat\Delta\right)(\hat C\hat c -\frac{1}{2})+\hbar\omega \hat D \hat d+\left( \frac{\hbar\omega_0}{2}+\rho \right)\1 .
\end{equation}

Following the general procedures used for the pseudo-fermions and pseudo-bosons operators  {in \cite{bagbook}, we can construct the eigenvectors of $H$ and $H^\dagger$ in the framework of} deformed canonical commutation relations and canonical anti-commutation relations. We know  that two non-zero vectors $\hat\varphi_0$ and $\hat\psi_0$ do exist in $\Hil_b$ such that, if $\hat\eta_0$ and $\hat\mu_0$ are two vectors of the fermionic Hilbert space $\Hil_f$ annihilated respectively by $\hat c$ and $\hat C^\dagger$, we have
\begin{equation}
\left(\hat d\otimes\hat \1_f\right)\hat\Phi_{0,0}=\left(\hat \1_b\otimes\hat c\right)\hat\Phi_{0,0}=0,
\label{26}
\end{equation}
as well as
\begin{equation}
\left(\hat D^\dagger\otimes\hat \1_f\right)\hat\Psi_{0,0}=\left(\hat \1_b\otimes\hat C^\dagger\right)\hat\Psi_{0,0}=0 \, ,
\label{27}
\end{equation}
where $\hat\Phi_{0,0}:=\hat\varphi_0\otimes\hat\eta_0$ and $\hat\Psi_{0,0}:=\hat\psi_0\otimes\hat\mu_0$.
As already pointed out in \cite{BL13,BGVo}, it is convenient to assume that $\hat\varphi_0$ and $\hat\psi_0$ belong to a dense domain $\Dbos$ of $\Hil_b$, which is {left} stable under the action of $d$, $D$, and their adjoint. {As for $\hat\eta_0$ and $\hat\mu_0$, these vectors surely exist in $\Hil_f$ and belong to the domain of all the (pseudo-fermionic) operators involved into the game, as one can easily deduce from the fact that $\Hil_f$ is a finite dimensional vector space}.
If such a $\Dbos$ exists, then we can use the two vacua $\hat\Phi_{0,0}$ and $\hat\Psi_{0,0}$ to construct two different sets of vectors, $\F_{\hat\Phi}:=\{\hat\Phi_{n,k}, \,n\geq0, \, k=0,1\}$ and
$\F_{\hat\Psi}:=\{\hat\Psi_{n,k}, \,n\geq0, \, k=0,1\}$, all belonging to $\Dbos\otimes\Hil_f$, as follows:
\begin{eqnarray}
\hat\Phi_{n,k}&=&\left(\frac{1}{\sqrt{n!}}\hat D^n\otimes \hat C^k\right)\hat\Phi_{0,0} \, ,
\nonumber \\
&=&\left(\frac{1}{\sqrt{n!}}\hat D^n\hat\varphi_0\right)\otimes\left(\hat C^k\hat\eta_0\right):=\hat\varphi_n\otimes\hat\eta_k \, \
\label{28}
\end{eqnarray}
and
\begin{eqnarray}
\hat\Psi_{n,k}&=&\left(\frac{1}{\sqrt{n!}}\hat {d^\dagger}^n\otimes \hat {c^\dagger}^k\right)\hat\Psi_{0,0}
\nonumber \\
&=&\left(\frac{1}{\sqrt{n!}}\hat {d^\dagger}^n\hat\psi_0\right)\otimes\left(\hat {c^\dagger}^k\hat\mu_0\right):=\hat\psi_n\otimes\hat\mu_k \, ,
\label{29}
\end{eqnarray}
with obvious notations, where $n=0,1,2,\ldots$ and $k=0,1$. It is now easy to check that
\begin{equation}
H\hat\Phi_{n,k}=E_{n,k}\hat\Phi_{n,k},\qquad H^\dagger\hat\Psi_{n,k}=E_{n,k}^*\hat\Psi_{n,k} \, ,
\label{210}
\end{equation}
where the eigenvalues are
\begin{equation}
E_{n,k}=\hbar\omega n+\frac{\hbar\omega_0}{2}+\rho+\left[\hbar \omega -\left(\delta^2+4|\epsilon|^2(n+k)\right)^{1/2} \right]\left(k-\frac{1}{2}\right).
\label{eigenvalues}
\end{equation}
Also, if the normalization of $\hat\Phi_{0,0}$ and $\hat\Psi_{0,0}$ is chosen in such a way that $\left<\hat\Phi_{0,0},\hat\Psi_{0,0}\right>=1$, then
\begin{equation}
\left<\hat\Phi_{n,k},\hat\Psi_{m,l}\right>=\left<\hat\varphi_{n},\hat\psi_{m}\right>_{\Hil_b}\left<\hat\eta_{k},\hat\mu_{l}\right>_{\Hil_f}=\delta_{n,m}\delta_{l,k} \, .
\label{biorth}
\end{equation}
Here $\left<.,.\right>_{\Hil_b}$ and $\left<.,.\right>_{\Hil_f}$ are respectively the scalar products in $\Hil_b$ and in $\Hil_f$.

\section{Exceptional Points formation}
\label{sec:EPform}
 {In this section we investigate the formation of exceptional points in our extended Jaynes-Cummings Hamiltonian.
It is convenient to introduce the analogous pseudo-structures given by the conditions (\ref{26}-\ref{29}) for the non diagonal form (\ref{31}) of our Hamiltonian. As before, we can also define the two  set of vectors,} $\F_{\Phi}:=\{\Phi_{n,k}, \,n\geq0, \, k=0,1\}$ and
$\F_{\Psi}:=\{\Psi_{n,k}, \,n\geq0, \, k=0,1\}$, all belonging to $\Dbos\otimes\Hil_f$, as follows:
\begin{eqnarray*}
\Phi_{n,k}=\left(\frac{1}{\sqrt{n!}} D^n\otimes  C^k\right)\Phi_{0,0}
\nonumber &=&\left(\frac{1}{\sqrt{n!}} D^n\varphi_0\right)\otimes\left( C^k\eta_0\right)=:\varphi_n\otimes\eta_k , \\
\Psi_{n,k}=\left(\frac{1}{\sqrt{n!}} {d^\dagger}^n\otimes  {c^\dagger}^k\right)\Psi_{0,0}&=&\left(\frac{1}{\sqrt{n!}} {d^\dagger}^n\psi_0\right)\otimes\left( {c^\dagger}^k\mu_0\right)=:\psi_n\otimes\mu_k \, ,
\label{pseudooriginal}
\end{eqnarray*}
and, as usual,
\begin{eqnarray}
\left( d\otimes\hat \1_f\right)\Phi_{0,0}=\left( \1_b\otimes c\right)\Phi_{0,0}=0,\\
\left( D^\dagger\otimes \1_f\right)\Psi_{0,0}=\left( \1_b\otimes C^\dagger\right)\Psi_{0,0}=0,
\label{pseudozero}
\end{eqnarray}
and
\begin{equation}
\left<\Phi_{n,k},\Psi_{m,l}\right>=\delta_{n,m}\delta_{l,k}.
\label{212bis}
\end{equation}
It is easy to check that, in terms of the vectors given above, the eigenvalues of $H$ and $H^\dag$ can be rewritten in order to satisfy the following conditions:

\begin{eqnarray}
H\left(\Phi_{n-1,1}+\lambda^{\pm}_{n}\Phi_{n,0} \right)&=&  E^{\pm}_{n}\left(\Phi_{n-1,1}+\lambda^{\pm}_{n}\Phi_{n,0} \right),\\
H^\dag\left(\Psi_{n-1,1}+\xi^{\pm}_{n}\Psi_{n,0} \right)&=&  E^{\pm *}_{n}\left(\Psi_{n-1,1}+\xi^{\pm}_{n}\Psi_{n,0} \right),
\label{210_bis}
\end{eqnarray}
where

$$ E^{\pm}_{n}=\hbar\omega (n-\frac{1}{2})+\frac{\hbar\omega_0}{2}+\rho\pm\frac{\left(\delta^2+4|\epsilon|^2n\right)^{1/2}}{2},$$
and

$$\lambda^{\pm}_{n}=\frac{-\delta\pm(\delta^2+4|\epsilon|^2n)^{1/2}}{2\epsilon\sqrt{n}}, \xi^{\pm}_{n}=\frac{-\delta^*\pm\left( (\delta^*)^2+4|\epsilon|^2n\right)^{1/2}}{2\epsilon\sqrt{n}}.$$\\
 {For each $n$, we have $ E^+_n=E_{n,0}$ and $ E^-_n=E_{n-1,1}$. It thus follows from \eqref{210} and \eqref{210_bis} that}
\begin{eqnarray*}
\hat \Phi_{n,0}&=&g^{n,0}_{\hat \Phi}\left(\Phi_{n-1,1}+\lambda^{+}_{n}\Phi_{n,0} \right),
\hat \Psi_{n,0}=g^{n,0}_{\hat \Psi}\left(\Psi_{n-1,1}+\xi^{+}_{n}\Psi_{n,0} \right),\\
\hat \Phi_{n-1,1}&=&g^{n-1,1}_{\hat \Phi}\left(\Phi_{n-1,1}+\lambda^{-}_{n}\Phi_{n,0} \right),
\hat \Psi_{n-1,1}=g^{n-1,1}_{\hat \Psi}\left(\Psi_{n-1,1}+\xi^{-}_{n}\Psi_{n,0} \right),
\end{eqnarray*}
being  {$g_{\Phi},g_{\Psi}$ appropriate normalization constants given by the bi-orthogonality conditions \eqref{biorth}.}

\begin{figure}
\begin{center}
\subfigure[$$]{\includegraphics[width=10cm]{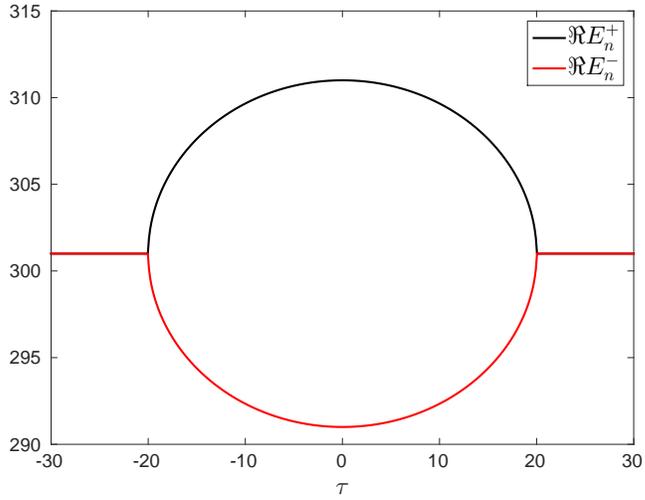}}
\subfigure[$$]{\includegraphics[width=10cm]{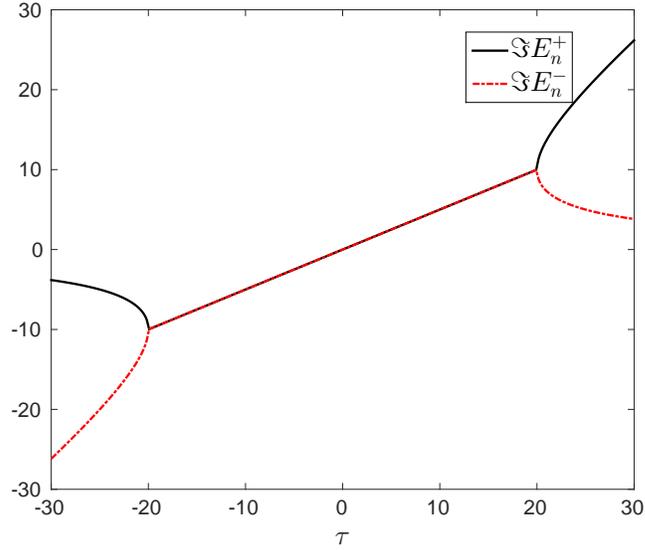}}
\caption{. Real ($\textbf{a}$) and imaginary ($\textbf{b)}$ parts of the eigenvalues $E_n^{\pm}$ for $n=100$ as function of the parameter $\tau=-i(\hbar\omega_0-\hbar\omega)$. Other parameters are $\epsilon=1,\rho=1,\omega=3$;  {units are such that $\hbar =1$.} At the EPs $\tau=\pm20$ eigenvalues coalesce so that $E_n^+=E_n^-$.}
\label{tauvar}
\end{center}
\end{figure}

If we now consider the case in which $\delta^2+4|\epsilon|^2n=0$, then
$ E_n^+= E_n^-,\lambda_n^+=\lambda_n^-$.  {Therefore, the couples of vectors} $\hat \Phi_{n,0},
\hat \Phi_{n-1,1}$ and $\hat \Psi_{n,0},
\hat \Psi_{n-1,1}$ being proportional, are linearly dependent. This condition  {implies} that $\delta$ is purely imaginary, so that $\hbar(\omega_0-\omega)=i\tau$ with $\tau=\tau^{\pm}=\pm2|\epsilon|\sqrt{n}$. Varying $\tau$ leads to the situation shown in Fig.\ref{tauvar} for $n=100$. For $\tau<\tau^-$ and $\tau>\tau^+$,
the eigenvalues $E_n^{+},E_n^-$ relative to $n$-excitation space have same real parts and different imaginary parts. For $\tau^-\leq\tau\leq\tau^{+}$, the eigenvalues have different real parts and the same imaginary parts. At $\tau=\tau^{\pm}$ the eigenvalues coalesce to the value $\hbar\omega(n-\frac{1}{2})+\frac{\hbar\omega_0}{2}+\rho$; also, due to the presence of two branch points in $\tau^{\pm}$ for $E_n^{\pm}(\tau)$, we obtain that encircling the points $\tau^{\pm}$ in the complex plane interchanges the two eigenvalues. In fact considering
an arbitrary closed loop $s^{\pm}(\theta)=\tau^{\pm}+re^{i\theta}$ around $\tau^{\pm}$ leads to $E_n^{\pm}(s(0))=E_n^{\mp}(s(2\pi))$.

\begin{figure}
\begin{center}
\subfigure[$$]{\includegraphics[width=10cm]{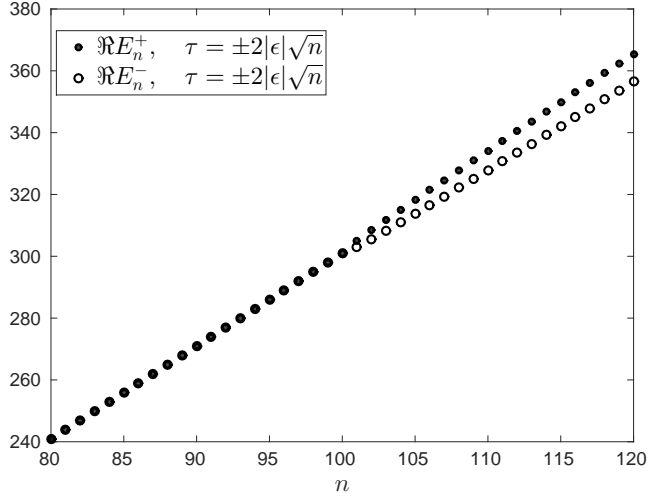}}
\subfigure[$$]{\includegraphics[width=10cm]{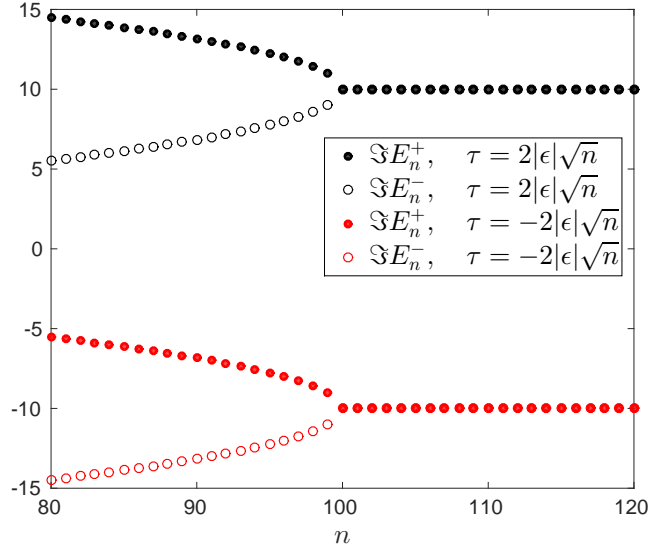}}
\caption{. Real ($\textbf{a}$) and imaginary ($\textbf{b)}$ parts of the eigenvalues $E_n^{\pm}$ as function of the parameter $n$ for $\tau=i(\hbar\omega_0-\hbar\omega)$ (black lines) and $\tau=-i(\hbar\omega_0-\hbar\omega)$ (red lines). Other parameters are $\epsilon=1,\rho=1,\omega=3$;  {units are such that $\hbar =1$. At $n=100$ the EPs are formed, and} eigenvalues coalesce so that $E_n^+=E_n^-$. }
\label{nvar}
\end{center}
\end{figure}

It is worth noting that at $\tau=\tau^{\pm}$ the condition $\lambda_n^+=\lambda_n^-=-\frac{\delta}{2|\epsilon|\sqrt{n}}$ leads to the vanishing of the scalar products $\left<\hat\Phi_{n,0},\hat\Psi_{n,0}\right>$ and $\left<\hat\Phi_{n-1,1},\hat\Psi_{n-1,1}\right>$. In fact
\begin{equation}
\left<\hat\Phi_{n,0},\hat\Psi_{n,0}\right>=
g^{n,0*}_{\hat \Phi}g^{n,0}_{\hat \Psi}\left( \left<\Phi_{n-1,1},\Psi_{n-1,1}\right>-\frac{\tau^2}{4|	\epsilon|^2n}\left<\Phi_{n,0},\Psi_{n,0}\right>\right)=0,
\end{equation}
because  $\tau^{\pm}=\pm2|\epsilon|\sqrt{n}$. Analogously
$\left<\hat\Phi_{n-1,1},\hat\Psi_{n-1,1}\right>$=0.
These conditions are typical of the EPs formation at $\tau^{\pm}$, \cite{Heiss13}. Fig.\ref{nvar} shows that only for $n=100$, for our particular choice of parameters, $E_n^{+}=E_n^{-}$, so that the related eigenstates coalesce.
More important,  our results show that at the EPs the pseudo-structure in terms of pseudo-fermions and pseudo-bosons operator is no more valid. In fact it has been shown that, in presence of an EP,  the coefficients defining the operators $c$ and $C$, see \eqref{add1}, cannot satisfy the necessary condition (\ref{PFexistence}) {(see \cite{FBgarg})}.

\begin{figure}
\begin{center}
\hspace*{-0cm}\subfigure[$$]{\includegraphics[width=9cm]{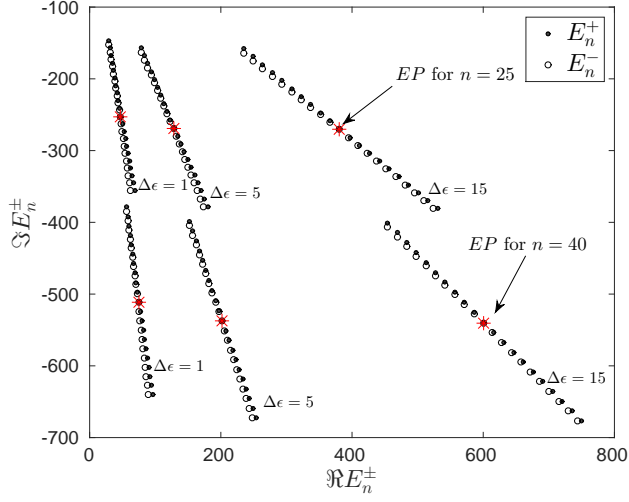}}
\hspace*{-0cm}\subfigure[$$]{\includegraphics[width=9cm]{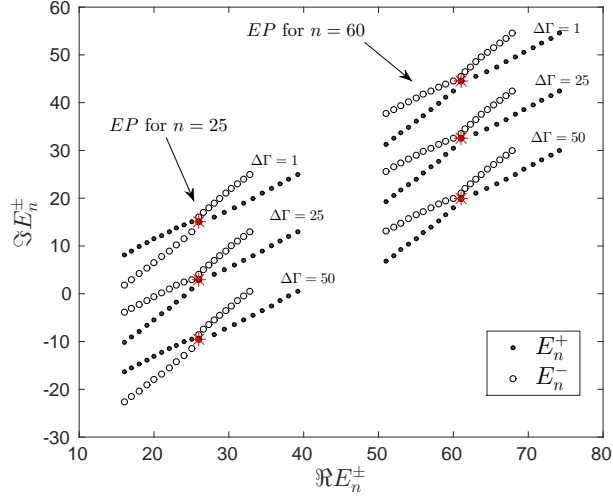}}
\caption{Eigenvalues $E_n^{\pm}$ in the complex plane for various  {values of} $n$. In \textbf{a)} eigenvalues are obtained by taking  $\nu_0=1,\epsilon_1=0.5,\Gamma_1=0,\Gamma_2=1$,
$\omega_0=\sqrt{4+(\Delta \epsilon+i)^2}$, $\rho=0.5(1+\Delta\epsilon-i+\sqrt{4+(\Delta \epsilon+i)^2})$ and $\omega=\sqrt{4+(\Delta \epsilon+i)^2}-2i\sqrt{\tilde n}$, for $\tilde n=25$ and $\tilde n=40$, $\epsilon_2$ is varied according to the chosen value of $\Delta\epsilon$.
In \textbf{b)} eigenvalues are obtained by taking  $\epsilon_1=\epsilon_2=0.5,\Gamma_1=0,\omega=1+i$,
$\nu_0=0.5\sqrt{(1+i-2i\sqrt{\tilde n})^2-(i\Delta\Gamma^2)},\omega_0=1+3i\sqrt{\tilde n},\rho=0.5(2-i\Delta\Gamma+3i\sqrt{n})$, for $\tilde n=25$ and $\tilde n=60$,
$\Gamma_2$ is varied according to the chosen value of $\Delta\Gamma$. At the EPs, the eigenvalues coalesce, marked  {in the plots} by the red $*$}
\label{nvarDepsilon}
\end{center}
\end{figure}

\begin{figure}
\begin{center}
\includegraphics[width=9cm]{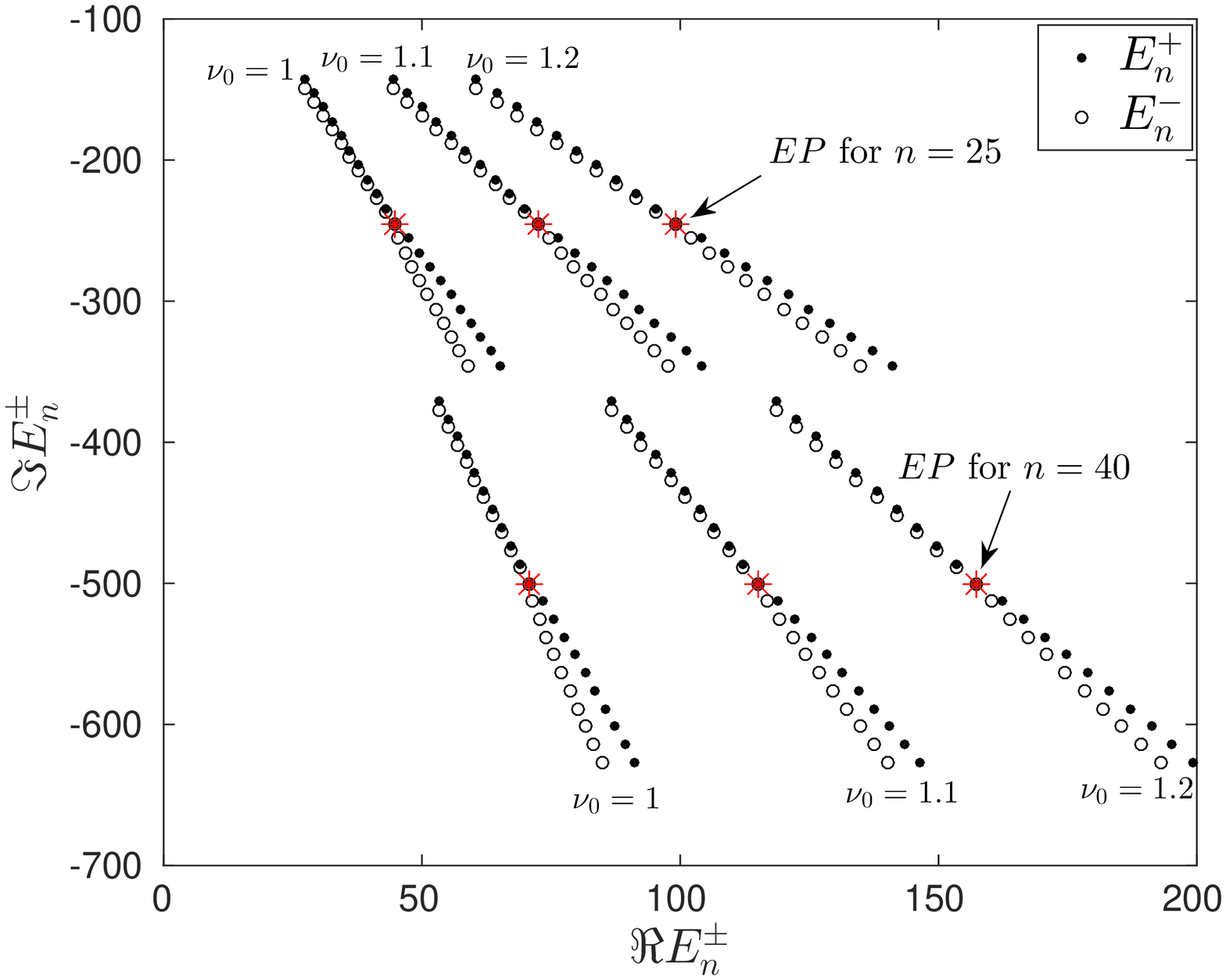}
\caption{Eigenvalues $E_n^{\pm}$ in the complex plane for various  {values of} $\nu_0$ and $n$. Parameters are  $\epsilon_1=\epsilon_2=0.5,\Gamma_1=0,\Gamma_2=1$,$\omega=1+1i,\omega_0=$, $\rho=0.5(1-i1+\omega_0)$, $\omega=\sqrt{4+(\Delta \epsilon+i1)^2}-2i\sqrt{\tilde n}$, for $\tilde n=25$ and $\tilde n=40$. At the EPs the eigenvalues coalesce, marked  {in the plot} by the red $*$}
\label{nvarNi0}
\end{center}
\end{figure}

Notice that through \eqref{relomega0}, we obtain that the EPs formation is compatible only with the following choices of $\omega$
\begin{equation}
\hbar\omega^{\pm}=\pm\sqrt{4\nu_0^2+(-\Delta\epsilon +i\Delta \Gamma)^2}-i\tau,
\end{equation}
where it is evident that EPs form only for {an} appropriate choice of $\nu_0$ and of the relative  {differences} $\Delta\epsilon,\Delta\Gamma$ of the parameters $\epsilon_{1,2}$ and $\Gamma_{1,2}$ in
\eqref{31},  {which is related to} $\omega$ and $\omega_0$.
It must be emphasized that  {by} manipulating parameters present in the Hamiltonian \eqref{31},  it is possible to change the position of the EPs in the complex plane. In Figs.\ref{nvarDepsilon}-\ref{nvarNi0} we show the eigenvalues $E_n^{\pm}$ in the complex plane by varying respectively $\Delta\epsilon,\Delta\Gamma$
and $\nu_0$. This opens up the prospects of a kind of engineering of EPs in order to exploit their impact on the dynamics of the physical system in which they appear. For example, this can be obtained by  {appropriately} changing the decay rates $\Gamma_1$ and $\Gamma_2$  (see \cite{LinGar06} and references therein).


\section{Conclusions and perspectives}
 {In this paper we have considered the formation of exceptional points in a non-Hermitian Jaynes-Cummings Hamiltonian, that generalizes the Hamiltonian of a two-level atom interacting with a single cavity field mode to the case in which dissipation and decay are phenomenologically included.

The results obtained in this paper show the exceptional points identified in the non-Hermitian Jaynes-Cummings Hamiltonian (\ref{31}), have the same structure obtained in \cite{MullRott08,EleRott14}.  

From a mathematical point of view, this is due to the fact that, for a two-level system, the eigenvalues contain in their mathematical expression a square root as that of a second-degree algebraic equation.
The collapse of the eigenvalues and the formation of EPs depend on the vanishing of the square-root argument for specific values of the physical parameters involved. On the other hand,} the interchange properties of the eigenvalues when EPs are encircled can be interpreted as an effect due to the branch points of the square root when analyzed as a function in the complex plane. Also, encircling EPs causes the switch of  {the} eigenfunctions, showing that their relative phases are not rigid. From a physical point of view, this behavior can be interpreted as a manifestation of the capability of the system to align itself with the environment to which it is coupled (see also \cite{EleRott14}).

The main novelties introduced in this paper concern  {with the analysis of the exceptional points for the spectrum of a non-Hermitian Jaynes-Cummings Hamiltonian expressed in terms of a mixture of pseudo-fermions and pseudo-bosons (previous analysis were focused on EPs in non-Hermitian pseudo-fermionic operators, \cite{FBgarg}). Thus, the application range} of this theory on pseudo-particles is here extended.
 {We wish to stress that the existence of a pseudo-structure is deeply
related to the existence of EPs: in fact, as we have shown in Section \ref{sec:EPform}, the spontaneous generation of the EPs implies that the biorthogonality condition \eqref{biorth}, which characterizes the pseudo-structure we have introduced, is no more satisfied.} This is expected, since a pseudofermionic or pseudobosonic
structure is intrinsically connected with the existence of
non coincident eigenvalues.
Moreover, the {\em deformed} Jaynes-Cummings model analyzed in \cite{BLPRS15} and in this paper, could be used to further investigate the role played by EPs on interaction between atomic systems and the electromagnetic field, including damping or amplifying processes, which is of fundamental importance in quantum optics.

\label{sec:Concl}

\end{document}